\documentclass[a4paper,twoside,11pt]{article}

\usepackage[english]{babel}
\usepackage[T1]{fontenc}
\usepackage{fancyhdr}
\usepackage{abstract}
\usepackage{graphicx}
\usepackage{amsmath} 
\usepackage[utf8]{inputenc}

\usepackage{html,makeidx}

\title{Thermal pseudo stress in rotating matter}
\author{David Jonsson, Sweden\thanks{phone \htmladdnormallink{+46703000370}{callto:+46703000370}, email: \htmladdnormallink{david@djk.se}{mailto:david@djk.se}}}
\date{Dec 2010 and ongoing as \htmladdnormallink{arXiv:1012.1381}{http://arxiv.org/abs/1012.1381}}

\begin{document}
\pagestyle{empty}

\twocolumn[

\maketitle

\vspace{-5mm}

\begin{onecolabstract}

Thermal molecular motion in an accelerated coordinate system causes variety in pseudo forces and thus variety in pseudo stress in rotating matter. Averaged Eötvös effects and Coriolis accelerations on molecular level become noticeable. The effect is here derived for gas but does also exist in liquids, solid matter and plasma. In the atmosphere of Venus and Earth the pseudo shear stress is negligible and the vertical normal pseudo stress from Eötvös effects is lowered by 0.51 \% and 0.27 \% respectively. The shear stress component is not dependent on temperature and the normal stress component increases with the square root of the temperature.

\vspace{7mm}

\end{onecolabstract}
]

\saythanks

\thispagestyle{empty}



\section*{Introduction}

High wind speeds in upper layers of the atmosphere requires an explanation. On Venus the winds blow 60 times faster than Venus surface. On Saturn they blow 500 m/s faster than the planet. Even several stars like the Sun have the same effect where outer equatorial layers of the star rotate faster than matter closer to the axis. Something has to drive these motions since viscosity acts as an attenuator of the difference in flow speed.

\section*{Derivation of shear stress produced by differences in centrifugal acceleration on rotating gas considering thermal molecular motion}

A small part of a gas is seen in a frame of reference fixed to the rotating system. A two dimensional cross section in the plane of rotation is described in figure 1. The angular momentum changes of the gas molecules are analyzed. Angular momentum change per time and volume is equal to shear stress. 

Usually fluid mechanics uses the average molecular motion in describing fluid velocity. Such a description is sometime too coarse and hides thermal pseudo stress. In order to find the Eötvös effects on the fluid components a more detailed description is needed.

In order to investigate the thermal pseudo stress the molecular motion is divided into six components: up, down, forward, backward, left and right. Molecular motion is assumed to be equally distributed in the three dimensions and in the six directions mentioned relative the mean flow of the fluid. Since the velocity terms are used in first and second order only the respective means speed of $\left< v_x \right>$ and $ v_{rms} $ instead of integrating the expressions from the velocity distribution. This makes the velocity description in the general case to increase from a three dimensional velocity field to a six dimensional field or since the thermal speed is considered the same in all directions the flow field becomes reduced to 4 dimensions. The worst complication is however not this additional thermal speed but the fact that the equations describing the flow, typically Navier Stokes equations, becomes more nonlinear than before since the stress term becomes dependent on acceleration of the flow.  

One case where the pseudo stress in noticeable is in the case of rotation. Centrifugal acceleration is considered to be a fictitious acceleration related to a rotating coordinate system. It is directed inwards. In an inertial coordinate system the centrifugal force is no essential force but an inertial reaction to the centripetal force.

The effect in this article is derived for gases. For liquid and solid matter the effect is harder to determine since the amplitude of molecular or atomic thermal vibration is less known. Also the potential functions of molecules becomes more important in liquid and solid matter. The approach used here is spring models in describing the potential function of displaced liquid and solid molecules.

\subsection*{Parameters used in derivation of pseudo stress in the case of centrifugal acceleration on a planets surface}
Molecular mean speed in one direction: $\left< v_x \right>$ \\
Gravitational acceleration: g \\
The molecules distance to the rotational axis, typically a planets radius: r \\
The systems angular velocity, for a planet the sidereal day speed plus wind speed: $\Omega $

\begin{figure*}[htb]
\centering
\includegraphics[clip,trim=42 415 42 42,scale=0.7]{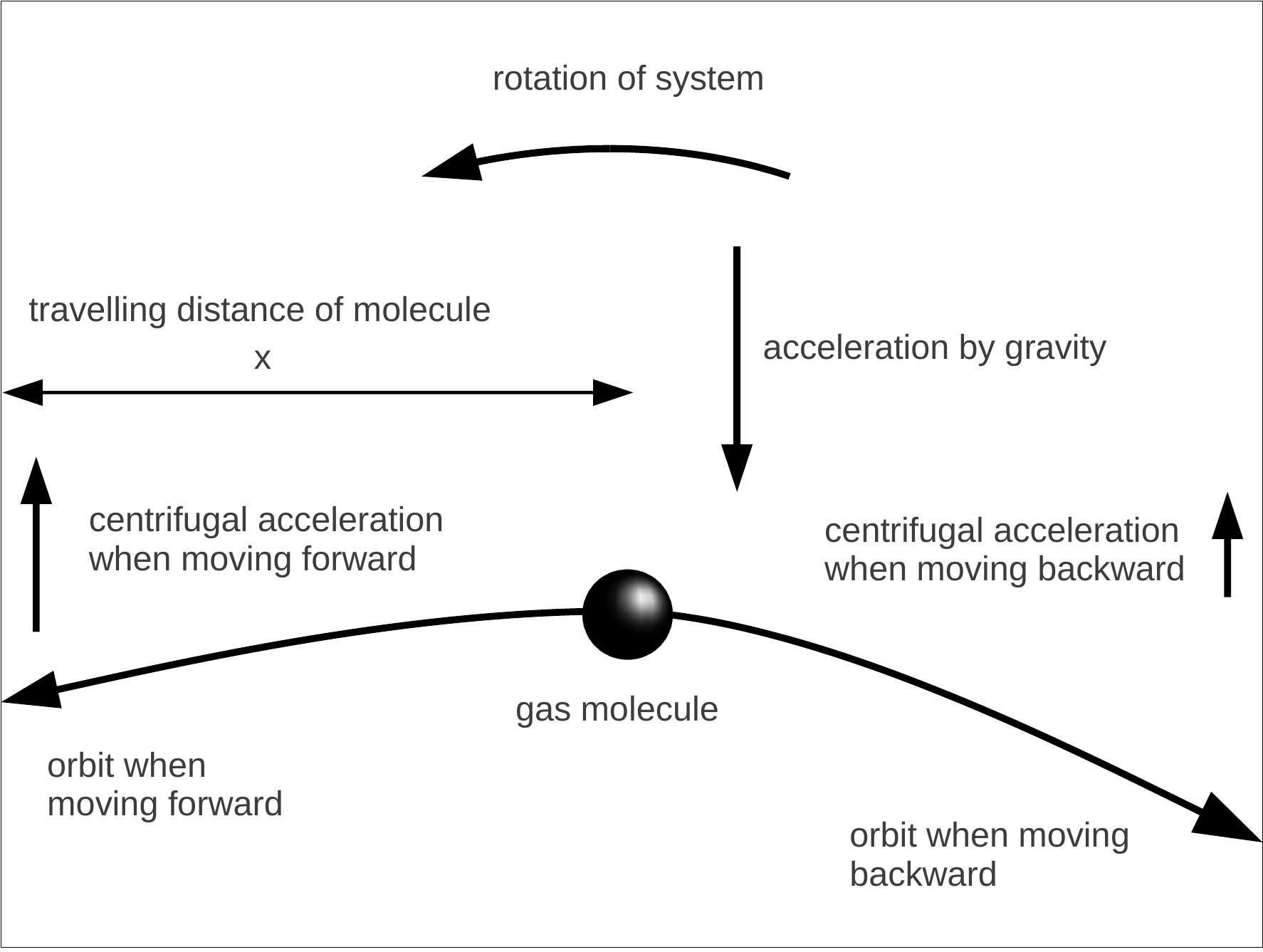}
\caption{A gas molecule in a rotating system}
\label{Molecule}
\end{figure*}

\subsection*{Derivation}
Motion in and out and motion up and down does not affect the angular momentum of the moving particle. The particle pair moving in and out is affected by gravity but their change in angular momentum balances each other and has no net effect. The motion up and down is affected by changes in the linear momentum and causes the for gases well-known adiabatic heat gradient. 

Particle motion forward in the direction has higher centrifugal acceleration compared to a particle moving backwards as is seen in figure~\ref{Molecule}. This causes a net difference in vertical speed between the molecule pair equivalent of a shear flow or a shear stress if seen as a continuum with a viscosity.

Shear stress can be calculated from shear flow if the viscosity is known according to Newtons postulate 
\begin{equation}
\label{eqn_5}
\tau= \mu \frac{\partial v_y }{ \partial x }
\end{equation}
where $\mu$ is viscosity and $\partial v_y / \partial x $ is shear flow ($\partial v_y$ is difference in vertical speed and $\partial x$ is difference in horizontal direction). Based on a particle pair in figure~\ref{Molecule} it is easy to calculate the shear flow and the stress. The speed difference is determined as the speed difference between a forward moving molecule and a backward moving molecule. $v_{xf}$  is the horizontal speed component for the forward moving particle and $v_{xb}$ is the horizontal speed component for the particle moving backwards.\\
$\partial x = (v_{xb} - v_{xf}) \partial t = \\
= ((\Omega r - \left< v_x \right>) - (\Omega r + \left< v_x \right>)) \partial t = \\
= -2 \left< v_x \right> \partial t $ \\
The accelerating force of gravity is g. The centrifugal acceleration is\\
$a = v^2/r $\\
and for the forward moving molecule acceleration is \\
$a_{cf} = (r \Omega + \left< v_x \right>)^2/r $ \\
The centrifugal acceleration for backwards moving molecule it is \\
$a_{cb} = (r \Omega - \left< v_x \right>)^2/r $ \\
The vertical speed differential thus becomes \\
$\partial v_y = \partial v_{yb} - \partial v_{yf} = \\
=(a_{cb} - g) \partial t - (a_{cf} - g) \partial t = \\
= (\Omega r- \left< v_x \right>)^2/r - (\Omega r+ \left< v_x \right>)^2/r) \partial t =\\
= -4\Omega \left< v_x \right>\partial t $ \\
leading to the expression of the stress (\ref{eqn_5}) to become 
\begin{equation}
\label{tauct}
\tau= \mu \frac{\partial v_y }{ \partial x} = \mu \frac{-4\Omega \left< v_x \right>\partial t }{-2 \left< v_x \right> \partial t}= 2\mu\Omega 
\end{equation}

\section*{Comparing the shear stress with observed shear flows}
Calculations based on Newtons postulate could be done for Venus and Earth but physical conditions for other planets and stars appears insufficiently known to allow a determination of the effect. In the extreme heat of the interior of giant gas planets the effect becomes noticeable and will likely lead to a change of the assessment of the physical properties of such planets.

\subsection*{The centrifugal thermal shear stress in the Earth atmosphere}
A calculation of the centrifugal thermal shear stress is calculated for STP at the Earth surface based on the following data. \\
Air viscosity at STP of $\mu=1.8 \cdot 10^{-5} Nsm^{-2}$ \\
Earth angular velocity $\Omega = 73 \mu radians/s $

The shear flow is \\
$ \frac{\partial v_y }{ \partial x} = 2\Omega = 9.2 \cdot 10^{-4} s^{-1}$ \\
In case of equilibrium
\begin{equation}
\label{equilibrium}
\frac{\partial v_y }{ \partial x} = \frac{\partial v_x }{ \partial y}
\end{equation}
The wind speed change per altitude on Earth is 0.92 m/s per kilometer. This appears to be too low to be distinguishable from the measured wind speed. The shear stress is $\tau = 1.7 nN/m^2$ \newline 

\subsection*{The centrifugal thermal shear stress in the Venusian atmosphere}
Observations of gas composition, temperature, pressure and shear flow exists for different altitudes. 
Gas is almost all carbon dioxide. \\
Venus atmosphere angular velocity \\
 $\Omega = 0.3 \mu radians/s$ + wind speed up to 3.5 $\mu$radians/s at 65 km height \\
Observed troposphere shear flow is $\partial u / \partial y = 0.003 s^-1$ according to Svedhem et al. \cite{Svedhem}.
The equilibrium condition \ref{equilibrium} leads to a contribution to the shear flow of 2 $\Omega$ which lies in the range from 0.6 to 7 $\mu s^{-1}$ or a factor 430 to 5000 lower than observed shear flow.

\subsection*{The gas centrifuge}
Instabilities have been observed in gas centrifuges past certain speeds. Solid body rotation is commonly assumed for the gas inside a gas centrifuge. If pseudo stress is present in the gas in a centrifuge it might not rotate as a solid body. Shear flow could be so strong that Kelvin-Helmholtz instabilities might occur. This could explain the instabilities at high centrifuge speed.

\section*{Laboratory investigation of the effect}
Both desktop centrifuges and gas centrifuges exist with centrifugal acceleration of a million g. The centrifugal thermal shear stress in such centrifuges could be noticeable especially since $\Omega$ is so much higher. Centrifuges seem like suitable equipment to experimentally test the shear stress effect. Another way could be to use small vibrating balls in rotating containers to simulate gas molecules.

\section*{Effects on normal stress, net central acceleration g, pressure, adiabatic temperature gradient and scale height}
The adiabatic temperature gradient (sometimes called lapse rate) is a related effect. It is well investigated and does not need any gas kinetic theory to be determined or explained but can also be determined and explained with kinetic gas theory. Scale height is similar to adiabatic heat gradient. The adiabatic heat gradient, scale height and the centrifugal thermal shear stress are similar in the sense that both are caused by pseudo forces and molecular thermal motion. The net acceleration of particles are affected by variety in centrifugal acceleration. This is the Eötvös effect. It can also be seen as a variety in normal stress. Despite the net arithmetic mean velocity of thermal motion being zero the alteration in centrifugal acceleration is not zero which will be shown below.

Adiabatic heat gradient is defined as 
\begin{equation}
\label{heatgradient}
{\gamma} = \frac{g}{ c_p}
\end{equation}
and scale height as 
\begin{equation}
\label{scaleheight}
H = \frac{kT}{Mg}
\end{equation}
where \emph{g} in (\ref{heatgradient}) and (\ref{scaleheight}) is defined as
\begin{equation}
\label{g}
g = -\frac{GM}{r^2} + \frac{v^2}{r} 
\end{equation}
and commonly \emph{v} is taken as surface speed of the rotating body, typically a planet or star. This is not very exact. The speed of each molecule should be used and the mean should be taken of the net acceleration which is affected by the two tangential velocity components of \emph{v}. This simplifies to using the root mean square speed for forward, backward, left and right motion and their variety do not cancel each other.

\subsection*{Derivation of g for gases}

Two speeds are used in the derivation. The net flow mean speed, usually surface speed, $v_s $, and rms speed in specific direction  \\
$v_{rmsd}=\frac{v_{rms}}{\sqrt{3}}$ 

For the thermal forward motion:\\
$v_f^2 = (v_s + v_{rmsd})^2 $\\
For the thermal backward motion:\\
$v_b^2 = (v_s - v_{rmsd})^2 $\\
The mean speed squared for forward and backward moving molecules \\
$(v_f^2+v_b^2)/2 = ((v_s^2 + 2v_sv_{rmsd} + v_{rmsd}^2) +(v_s^2 - 2v_sv_{rmsd} + v_{rmsd}^2) )/2 = v_s^2 + v_{rmsd}^2 $\\

is not equal to $v_s^2$.

The sideways north south motion of a molecule is perpendicular to  $v_s^2$ and should thus be calculated as
$v_{sideways}^2= v_s^2 + v_{rmsd}^2$ \\

The up down motion of a molecule does not affect centrifugal acceleration. This means that only 2/3 of the mentioned directional motions of the molecule affect a calculation of g.

The net total mean squared speed contributing to the acceleration g becomes 
\begin{equation}
\label{vgt}
v^2 = v_s^2 + 2 v_{rmsd}^2 =  v_s^2 + \frac{2}{3}v_{rms}^2
\end{equation}
leading to the updated form of equation (\ref{g}) to become
\begin{equation}
\label{gt}
g = -\frac{GM}{r^2} + \frac{ v_s^2 + \frac{2}{3}v_{rms}^2}{r} 
\end{equation}
For the Earth this leads to the adiabatic heat gradient (\ref{heatgradient}) to become 0.27 \% lower. For Venus (\ref{heatgradient}) becomes 0.51 \% lower at the surface.

The hot interior of big planets like Jupiter and Saturn the thermal motion contribution to the scale height could be significant. Calculation of the effect in the solar photosphere lowers g by $10^{-7}$

The effect also affects pressure and normal stress in the atmosphere correspondingly. Normal stress in vertical direction is defined as the weight per area and weight is density times g and thus dependent on temperature. If atmospheric density on Earth is determined by weighing air then density will be determined 0.16 \% too low.

\section*{Effects in liquids and solids}
From a gas kinetic perspective it is not hard to imagine a similar stress effect in liquids and solids. Molecules and atoms in liquids and solids move in a similar way as in gases but less far. The adiabatic heat gradient is derived or measured for some liquids and some solids. For the Earth mantle the heat gradient it is 0.3 K/km. It could thus be motivated to investigate if the centrifugal thermal shear stress can be derived or measured for a liquid or solid. In a liquid it could lead to shear flows similar to what is observed in stars and in the oceans of the Earth. The Earth mantle could have shear stress and shear flow as well as the liquid outer core. In solid matter the centrifugal thermal shear stress would lead to a static deformation and a non-vertical principal stress. Stress is not always vertical in the Earth crust.

\section*{Determining the stress in liquids and solids}
Liquids and solids have similarities with gases in the sense that the constituent particles have thermal motion. While thermal energy in gases is almost entirely in the form of kinetic energy the case for solids and liquids is different. Here the thermal energy is oscillating between kinetic and potential energy. This leads to the average speed of the particles to be lower and thus any dependent stress to become lower as well. By using the simple spring model for liquids and solids the value for $v_{rms}$ is lowered by $2^{-1/2}$ and the arithmetic mean in one positive specific direction $\left< v_{psd} \right>$ is lowered by $ 2/\pi $.

With this physical model the Eötvös variations lead to the following expression for central acceleration (\ref{gt}) to become 
\begin{equation}
\label{gtls}
g = -\frac{GM}{r^2} + \frac{ v_s^2 + \frac{2}{6}v_{rms}^2}{r} 
\end{equation}

For water at normal temperature 20 degrees Celsius (\ref{gtls}) becomes  0.22\% lower than (\ref{g}). This seems much and is probably lowered by collective effects where several molecules are moving together and acting as a particle with a mass totaling the sum of the mass of all molecules that move together.

\section*{General flow description}
The Navier Stokes equations 
\begin{equation}
\label{ns}
\rho \frac{D \mathbf{v}}{D t} = \nabla \cdot\mathbf{\sigma} + \mathbf{f}
\end{equation}

are well established in describing flows in general. In order to adapt them for the described effect the stress tensor $\mathbf{\sigma}$ has to include the speed and acceleration dependent shear and normal stress components derived in this article. Alternatively the Navier Stokes equations could be adjusted to a multiple fluid (n-fluid) for each of the directions used in this derivation. 

\section*{Effects on ionized matter and plasma}
Since positive and negative particles in a plasma have very different mass they also have very different thermal speed. Since the stress effect increases with increasing thermal speed electrons in a plasma will be much more affected by the centrifugal thermal stress compared to the ions. In the case of normal stress this will lead to negative electrons being less attracted to a gravitational body and thus produce a positive electric field. An electric field of 100 V/m is observed in the atmosphere of Earth. 

The centrifugal thermal shear stress can likewise give rise to a magnetic field. It could play a role in the formation of magnetic fields of rotating planets, stars and other rotating plasmas. Metallic rotating axes do get magnetic over time.

\section*{Comparison with viscosity}
Viscosity is a gas kinetic effect where molecular motion causes energy and momentum exchange with surrounding molecules. The centrifugal thermal stress is different from viscosity. The stress is present even in rigid body rotation of gas which viscosity is not. The centrifugal thermal shear stress is most likely balanced by an opposing stress from the viscous shear flow leading to balanced stationary shear flows as the ones seen on planets and stars.

\section*{Approximations made}
The gas kinetic values could be more precisely determined from the actual ensemble of molecular motions. The use of the mean molecular mass and speed of a gas instead of the mass and speed for each component gas can be an approximation. Collective phenomena in gases could affect the stress. Internal forces between molecules in a gas like van der Waals forces could be taken into consideration to acquire a higher precision. 




\end{document}